\documentclass[aip,jcp,reprint,onecolumn,superscriptaddress]{revtex4}
\usepackage{amsmath}
\usepackage{mathtools}
\usepackage{amsbsy}
\usepackage{bm}
\usepackage{amssymb}
\usepackage{graphics}
\usepackage{graphicx}
\usepackage[colorlinks]{hyperref}
\usepackage{color}
\usepackage{subfig}
\usepackage{epstopdf}
\usepackage{esint}
\usepackage{braket}
\usepackage[ruled,vlined]{algorithm2e}
\usepackage{multirow}
\usepackage{pseudocode}
\allowdisplaybreaks

\def\denM{{\mathcal{D}}}

\def\den{{\rho_{\mathcal{D}}}}

\def\R{{\mathbb{R}}}

\def\bRJ{{\mathbf{R}_{J}}}
\def\bRJp{{\mathbf{R}_{J'}}}
\def\be{{\mathbf{x}}}

\def\sab{{\sigma \mkern-2mu_{\alpha \beta}}}
\def\grada{{\nabla \mkern-6mu_{x_{\alpha}}}}
\def\gradb{{\nabla \mkern-6mu_{x_{\beta}}}}
\def\gradya{{\nabla \mkern-6mu_{y_{\alpha}}}}
\def\gradyb{{\nabla \mkern-6mu_{y_{\beta}}}}
\def\gradx{{\bm{\nabla \mkern-6mu_{x}}}}
\def\R{{\mathbb{R}}}

\def \gammaJl{\gamma_{_{Jl}}}
\def\tchins{{\tilde{\chi}_{_{Jlm}}}}

\def\pchins{{\chi_{_{J'lm}}}}

\def\pchis{{\chi_{_{J'lm}}}}

\newcommand{\bR}{\mathbf{R}}
\newcommand{\bx}{\mathbf{x}}
\newcommand{\by}{\mathbf{y}}

\newcommand{\dx}{\,\text{d}\bx}
\newcommand{\dy}{\,\text{d}\by}

\newcommand{\Vnl}{V_{nl}}
\newcommand{\hatHs}{\hat{\mathbf{H}}_\mathbf{s}}

\begin{document}

\title{Real-space density kernel method for Kohn-Sham density functional theory calculations at high temperature}

\author{Qimen Xu}
\affiliation{College of Engineering, Georgia Institute of Technology, GA 30332, USA}
\author{Xin Jing}
\affiliation{College of Engineering, Georgia Institute of Technology, GA 30332, USA}
\author{Boqin Zhang}
\affiliation{College of Engineering, Georgia Institute of Technology, GA 30332, USA}
\author{John E. Pask}
\affiliation{Physics Division, Lawrence Livermore National Laboratory, Livermore, CA 94550, USA}
\author{Phanish Suryanarayana}
\email[Email: ]{phanish.suryanarayana@ce.gatech.edu}
\affiliation{College of Engineering, Georgia Institute of Technology, GA 30332, USA}
\date{\today}

%%%%%%%%%%%%%%%%%%%%%%%%%%%%%%%%%%%%%%%%%%%%%%%%%%%%%%%%%%%%%%%%%%%%%%%%%%%%%%%%%%%%%%%%%%%%%%%%%%%%%%%%%%%%%%%%%%%%%%%%%%%%%%%%%%%%%%%%%%%%%%%%%%%%%%%%%%%%%%%%%%%%%%%%%%%%%%%%
\begin{abstract}
Kohn-Sham density functional theory calculations using conventional diagonalization based methods become increasingly expensive as temperature increases due to the need to compute increasing numbers of partially occupied states. We present a density matrix based method for Kohn-Sham calculations at high temperature that eliminates the need for diagonalization entirely, thus reducing the cost of such calculations significantly. Specifically, we develop real-space expressions for the electron density, electronic free energy, Hellmann-Feynman forces, and Hellmann-Feynman stress tensor in terms of an orthonormal auxiliary orbital basis and its density kernel transform, the density kernel being the matrix representation of the density operator in the auxiliary basis. Using Chebyshev filtering to generate the auxiliary basis, we next develop an approach akin to Clenshaw-Curtis spectral quadrature to calculate the individual columns of the density kernel based on the Fermi operator expansion in Chebyshev polynomials; and employ a similar approach to evaluate band structure and entropic energy components. We implement the proposed formulation in the SPARC electronic structure code, using which we show systematic convergence of the aforementioned quantities to exact diagonalization results, and obtain significant speedups relative to conventional diagonalization based methods. Finally, we employ the new method to compute the self-diffusion coefficient and viscosity of aluminum at 116,045 K from Kohn-Sham quantum molecular dynamics, where we find agreement with previous more approximate orbital-free density functional methods.
\end{abstract}

\maketitle

%%%%%%%%%%%%%%%%%%%%%%%%%%%%%%%%%%%%%%%%%%%%%%%%%%%%%%%%%%%%%%%%%%%%%%%%%%%%%%%%%%%%%%%%%%%%%%%%%%%%%%%%%%%%%%%%%%%%%%%%%%%%%%%%%%%%%%%%%%%%%%%%%%%%%%%%%%%%%%%%%%%%%%%%%%%%%%%%%
%%%%%%%%%%%%%%%%%%%%%%%%%%%%%%%%%%%%%%%%%%%%%%%%%%%%%%%%%%%%%%%%%%%%%%%%%%%%%%%%%%%%%%%%%%%%%%%%%%%%%%%%%%%%%%%%%%%%%%%%%%%%%%%%%%%%%%%%%%%%%%%%%%%%%%%%%%%%%%%%%%%%%%%%%%%%%%%%%
\section{Introduction}
Over the past few decades, Kohn-Sham density functional theory (DFT) \cite{Hohenberg, Kohn1965}  has established itself as a powerful framework for understanding and predicting a wide range of materials properties, from the first principles of quantum mechanics, without any empirical or ad hoc parameters. The ubiquitous use of  DFT is a consequence of its simplicity, generality, and high accuracy-to-cost ratio compared to other such first principles methods. However,  the solution of the underlying nonlinear eigenvalue problem for the Kohn-Sham orbitals remains a challenging task, with the computational cost and memory requirements scaling cubically and quadratically with system size, respectively \cite{Goedecker, Bowler2012, aarons2016perspective}. Moreover, the orthogonality constraint on the orbitals translates to significant global communications in parallel computing, limiting the minimum time to solution that can be attained. This can become particularly important for quantum molecular dynamics (QMD) simulations \cite{marx2009ab,kresse1993ab}, wherein tens or hundreds of thousands of such Kohn-Sham solutions may be required to complete a single simulation.

To overcome the cubic scaling bottleneck in DFT calculations, significant efforts have been directed towards the development of methods that scale linearly with system size (see, e.g., \cite{Goedecker, Bowler2012, aarons2016perspective} and references therein), both in computational cost and computer memory, which have culminated in a number of mature codes, e.g., \cite{SIESTA, Conquestref, ONETEP, FEMTECKref, MGMolref, BigDFTref,  FreeONref}. While major advances have been acheived, a number of challenges remain for linear scaling methods and their implementations. These include the need for additional computational parameters, which complicate use in practice; limitations of the underlying basis sets used for discretization; subtleties in choosing the numbers and/or centers of localized orbitals for different physical systems; scalability on parallel computing platforms due to complex communication patterns and challenges in load balancing; and calculation of accurate atomic forces and stresses as employed in structural relaxation and QMD  simulations \cite{aarons2016perspective, Bowler2012, RuiHinSky12}. Perhaps most importantly, the study of systems with partially occupied Kohn-Sham orbitals, as encountered in metallic systems or insulating systems at high temperature, remains particularly challenging.

Kohn-Sham QMD simulations at high temperature are employed in a variety of applications areas, such as warm dense matter and dense plasmas, as occur in fusion energy research and the inner regions of giant planets and stars \cite{gradesred2014, grabasben2011, renaudin2003aluminum, dharma2006static, ernstorfer2009formation, white2013orbital}. However, such calculations pose unique challenges in addition to those described above for ambient conditions. In particular, the number of orbitals that need to be computed increases with temperature, due to the increase in number of states that become partially occupied in the Fermi-Dirac distribution, which advances the onset of the cubic scaling bottleneck in diagonalization based methods, i.e., methods that calculate the Kohn-Sham orbitals. In addition, these orbitals are more diffuse, since higher-energy states become less localized. As a result, local-orbital based linear scaling methods suffer from large prefactors that increase rapidly with temperature. Consequently, Kohn-Sham QMD simulations at high temperature become impractical using either of these approaches.

There have been recent efforts to address the aforementioned challenges associated with high temperature calculations, including orbital-free molecular dynamics (OFMD) \cite{lamclegil2006}, in which a functional of the electron density is used to approximate the electronic kinetic energy; extended first principles molecular dynamics (ext-FPMD) \cite{zhang2016extended, blanchet2021extended}, in which planewaves are used to approximate the higher-energy orbitals; and finite-temperature potential functional theory (PFT) \cite{cangi2015efficient}, in which a coupling-constant formalism is used to develop an orbital-free approximation for the free energy functional. To address both scaling with system size and temperature, while retaining full Kohn-Sham accuracy for  insulators and metals alike, the  Spectral Quadrature (SQ) method for linear scaling calculations was recently developed \cite{suryanarayana2013spectral, pratapa2016spectral, suryanarayana2018sqdft}. In particular, the computational cost of the SQ method decreases with increasing temperature, a consequence of the faster decay of the density matrix, i.e., electronic interactions becoming more localized, and the increase in smoothness of the Fermi-Dirac distribution. In addition, the SQ method has excellent scaling in parallel computations since the communication pattern remains fixed and well localized to nearby processors throughout the computation. However, while the SQ method has proven highly accurate and efficient in applications reaching millions of kelvin \cite{wu2021development, zhang2019equation}, the associated prefactor becomes larger at less extreme temperatures, e.g., $\mathcal{O}$(10,000)--$\mathcal{O}$(250,000) kelvin, particularly when large numbers of grid points per atom are required.

In this work, we present a method, which we call SQ3, that is accurate and efficient at temperatures too high for efficient calculations using conventional diagonalization based methods but too low for efficient calculations using the SQ method. The combination of conventional diagonalization based methods at low temperatures, SQ3 at moderately high temperatures (e.g., $\mathcal{O}$(10,000)--$\mathcal{O}$(250,000) kelvin), and SQ at higher temperatures then brings accurate and efficient Kohn-Sham calculations to the full range of temperatures from ambient to millions of kelvin. As we detail below, the key idea of the method is to employ spectral quadrature to compute the density kernel --- i.e., density operator in a minimal orthonormal basis --- rather than required parts of the full density matrix --- i.e., density operator on the real-space grid --- as in the SQ method. In so doing, the need for diagonalization is eliminated and key operations are reduced to vectors and matrices of dimension equal to the number of occupied states rather than number of real-space grid points. We implement the method in the SPARC electronic structure code \cite{xu2021sparc, ghosh2017sparc2, ghosh2017sparc1}, where we find systematic convergence to exact diagonalization results and significant speedups relative to conventional diagonalization based methods.

The remainder of this paper is organized as follows. In Sections~\ref{Sec:RSDFT} and \ref{Sec:GS:Ds}, we present real-space expressions for the electron density, electronic free energy, Hellmann-Feynman atomic forces, and Hellmann-Feynman stress tensor in terms of the density operator and density kernel, respectively. In Section~\ref{Sec:SQ3}, we describe the formulation and implementation of the proposed SQ3 method, whose accuracy and efficiency are verified in Section~\ref{Sec:Results}. Finally, we provide concluding remarks in Section~\ref{Sec:Conclusions}. 

%%%%%%%%%%%%%%%%%%%%%%%%%%%%%%%%%%%%%%%%%%%%%%%%%%%%%%%%%%%%%%%%%%%%%%%%%%%%%%%%%%%%%%%%%%%%%%%%%%%%%%%%%%%%%%%%%%%%%%%%%%%%%%%%%%%%%%%%%%%%%%%%%%%%%%%%%%%%%%%%%%%%%%%%%%%%%%%%%
%%%%%%%%%%%%%%%%%%%%%%%%%%%%%%%%%%%%%%%%%%%%%%%%%%%%%%%%%%%%%%%%%%%%%%%%%%%%%%%%%%%%%%%%%%%%%%%%%%%%%%%%%%%%%%%%%%%%%%%%%%%%%%%%%%%%%%%%%%%%%%%%%%%%%%%%%%%%%%%%%%%%%%%%%%%%%%%%%

\section{Real-space DFT: density operator formulation} \label{Sec:RSDFT}
Consider a unit cell $\Omega$ with nuclei positioned at $\bR = \{\bR_1, \bR_2, \ldots, \bR_N \}$ and a total of $N_e$ electrons. Neglecting spin and Brillouin zone integration, 
the single-particle density operator $\mathcal{D}$ in Kohn-Sham DFT \cite{Kohn1965, hohenberg1964inhomogeneous} can be written as
%\begin{equation}
%\mathcal{D} = \sum_{i=1}^{N_s} g_i \Ket{\psi_i} \Bra{\psi_i} \,, \quad \mathcal{D}(\bx,\by) = \Bra{\bx} \denM \Ket{\by} = \sum_{i=1}^{N_s} g_i \psi_i(\bx) \psi_i(\by) \,, \label{Eqn:DensityMatrixDef}
%\end{equation}
\begin{equation}
\mathcal{D} = \sum_{i=1}^{N_s} g_i \Ket{\psi_i} \Bra{\psi_i}
\end{equation}
or, in real space,
\begin{equation}
\mathcal{D}(\bx,\by) = \Bra{\bx} \denM \Ket{\by} = \sum_{i=1}^{N_s} g_i \psi_i(\bx) \psi_i(\by) \,, 
\label{Eqn:DensityMatrixDef}
\end{equation}
where $N_s$ is the number of occupied states, and $\psi_i$ are the Kohn-Sham orbitals with energies $\lambda_i$ and occupations $g_i$ given by the Fermi-Dirac function $g$:
\begin{equation}
g_i = g(\lambda_i, \mu, \sigma) \equiv \left( 1 + \exp \left (\frac{\lambda_i - \mu}{\sigma} \right) \right)^{-1} \,.
\end{equation}
Above, $\sigma = k_B T$ is the smearing value, where $k_B$ is Boltzmann's constant and $T$ is the electronic temperature, and $\mu$ is the Fermi level determining the total number of electrons:
\begin{equation}
    2\text{Tr}(\mathcal{D}) \, = N_e\,, \label{Eqn:TraceD}
\end{equation}
where $\mathrm{Tr}(\cdot)$ denotes the trace of the operator.

The Kohn-Sham orbitals, energies, and occupations are solutions to the 
nonlinear eigenvalue problem
\begin{eqnarray}\label{Eqn:KSeqn}
\left( \mathcal{H} \equiv -\frac{1}{2} \nabla^2 +  V_{xc}[\rho_\denM] + \phi[\rho_\denM,\bR] + \Vnl[\bR]    \right)  \psi_i(\bx)  =  \lambda_i \psi_i(\bx) \,, 
%\quad n=1,2, \ldots, N_s \,,
\end{eqnarray}
where $\mathcal{H}$ denotes the Hamiltonian operator, $V_{xc}$ is the exchange-correlation potential, taken in the local density approximation (LDA) \cite{Kohn1965} in the present work, $\phi$ is the electrostatic potential \cite{pask2005, Phanish2010, Phanish2011}, $\Vnl$ is the nonlocal pseudopotential operator, and $\rho_\denM$ is the electron density:
\begin{equation} 
\den(\bx) = 2 \mathcal{D}(\bx,\bx) \,. \label{Eqn:rhoDensMat}
\end{equation}
The electrostatic potential  is the solution of the Poisson problem \cite{pask2005, ghosh2016higher, ghosh2017sparc2}
\begin{eqnarray}\label{eqn:poisson}
-\frac{1}{4\pi} \nabla^2 \phi(\bx,\bR) = \rho_{\denM}(\bx) + b(\bx,\bR) \,,
\end{eqnarray}
where $b$ is the total pseudocharge density. In addition, the nonlocal pseudopotential operator in Kleinman-Bylander form \cite{kleinman1982efficacious} is given by
\begin{eqnarray}\label{eqn:nonlocaloperator}
\Vnl  = \sum_{J} V_{nl,J} = \sum_{J} \sum_{lm} \gamma_{Jl} \Ket{\tilde{\chi}_{Jlm}} \Bra{ \tilde{\chi}_{Jlm}} \,,
\end{eqnarray}
where the summation index $J$ extends over all atoms in $\Omega$, $l$ and $m$ are azimuthal and magnetic quantum numbers, respectively, and $\tchins= \sum_{J'} \pchins $ are periodically extended nonlocal projectors, with $\pchins$ being the projectors of the $J'^{th}$ atom and $J'$ running over the $J^{th}$ atom and its periodic images.

The density operator can be written in terms of the Hamiltonian as
\begin{eqnarray}\label{Eq:denistyOperatorHam}
\mathcal{D} & = & g(\mathcal{H},\mu,\sigma) = \left( \mathcal{I} + \exp \left(\frac{\mathcal{H}-\mu \mathcal{I}}{\sigma} \right)   \right)^{-1} \,, \label{Eqn:DHRelation} 
\end{eqnarray} 
where $\mathcal{I}$ is the identity operator. Once the electronic ground state has been determined through the self-consistent solution of the above equation, the electronic free energy can be written as \cite{pratapa2016spectral, suryanarayana2018sqdft}
\begin{eqnarray} 
\mathcal{F} (\bR) & = & 2\text{Tr}(\mathcal{D} \mathcal{H})  + E_{xc}[\rho_{\mathcal{D}}] - \int_{\Omega} V_{xc}[\rho_{\mathcal{D}}(\bx)] \rho_{\mathcal{D}}(\bx) \, \mathrm{d\bx} + \frac{1}{2} \int_{\Omega} (b(\bx,\bR)-\rho_{\mathcal{D}}(\bx)) \phi(\bx,\bR) \, \mathrm{d\bx} \nonumber \\
& - & E_{\rm self}(\bR) + E_c(\bR) + 2 \sigma \text{Tr}\left( \mathcal{D} \log \mathcal{D} + (\mathcal{I}-\mathcal{D}) \log (\mathcal{I}-\mathcal{D}) \right) \,, \label{Eqn:ON:FreeEnergy}
\end{eqnarray}
where the first term is referred to as the band structure energy ($E_{band}$), the last term is the electronic entropy energy ($E_{ent}$) associated with partial occupations, $E_{\rm self}$ and $E_c$ are the self and overlap energy corrections, respectively, associated with the pseudocharges \cite{Suryanarayana2014524, ghosh2014higher}, and 
$E_{xc}$ is the exchange-correlation energy: 
\begin{eqnarray}
E_{xc} \left[\rho_{\mathcal{D}} \right] & = & \int_{\Omega} \varepsilon_{xc} \left[\rho_{\mathcal{D}}(\bx)\right] \rho_{\mathcal{D}}(\bx) \dx \,.
\end{eqnarray}
Here, $\varepsilon_{xc}$ is the sum of the exchange and correlation energy per particle of a uniform electron gas. 

The Hellman-Feynman  forces on the nuclei can be written as \cite{pratapa2016spectral, suryanarayana2018sqdft}
\begin{eqnarray}  
\mathbf{f}_{I} & = & \mathbf{f}_I^{l} + \mathbf{f}_I^{nl} \nonumber \\
& = & \sum_{I'} \int_{\Omega} \nabla b_{I'}(\bx,\bR_{I'}) \phi(\bx,\bR) \, \mathrm{d\bx} + \mathbf{f}_{sc,I} - 4 \text{Tr}\left(V_{nl,I} \nabla \mathcal{D} \right) \,, 
\label{Eqn:ON:forces}
\end{eqnarray}
where $\mathbf{f}_{sc,I} = - \frac{\partial (-E_{\rm self}(\bR) + E_c(\bR)) }{ \partial \bR_I }$ (see Ref.~\cite{xu2020matlab} for an explicit expression), $b_{I'}$ is the pseudocharge density of the ${I'}^{th}$ nucleus that generates potential $V_{I'}$, the summation index $I'$ runs over the $I^{th}$ atom and its periodic images, and $I$ extends over all atoms in $\Omega$. The first two terms together constitute the local component of the force ($\mathbf{f}_I^{l}$) and the last term is the nonlocal component of the force ($\mathbf{f}_I^{nl}$). 

The Hellmann-Feynman stress tensor can be written as \cite{sharma2020real}
\begin{equation}
\sab = \frac{1}{|\Omega|} \bigg[\sigma \mkern-2mu ^{I}_{\alpha \beta} + \sigma \mkern-2mu ^{II}_{\alpha \beta} + \sigma \mkern-2mu ^{III}_{\alpha\beta} + \sigma \mkern-2mu ^{IV}_{\alpha \beta}  \bigg] \,, \quad \alpha, \beta \in \{1, 2, 3\} \,, \label{Eqn:stressexpansion}
\end{equation}
where $|\Omega|$ is the volume of the unit cell, and $\sigma \mkern-2mu ^{I}_{\alpha \beta}$, $\sigma \mkern-2mu ^{II}_{\alpha \beta}$, $\sigma \mkern-2mu ^{III}_{\alpha \beta}$, and $\sigma \mkern-2mu ^{IV}_{\alpha \beta}$ are the contributions arising from the electronic kinetic energy, exchange-correlation energy $E_{xc}$, nonlocal pseudopotential energy $E_{nl}$, and the total electrostatic energy, respectively:  
\begin{eqnarray}
\sigma \mkern-2mu ^{I}_{\alpha \beta} & = &   
2  \int_\Omega  \bigg( \gradya \gradyb \mathcal{D}(\by,\bx)\bigg)\bigg|_{\by = \be} \dx \,, \label{Eq:sigmaI} \\
\sigma \mkern-2mu ^{II}_{\alpha \beta} & = & 
\delta_{\alpha \beta} \bigg( E_{xc}(\rho_\mathcal{D}) - \int_\Omega V_{xc}\big(\rho_\mathcal{D}(\bx) \big) \rho_\mathcal{D}(\bx) \dx \bigg)  \,, \label{Eq:sigmaII} \\
\sigma \mkern-2mu ^{III}_{\alpha \beta} & = & - 2\, \delta_{\alpha \beta} \sum_J \sum_{lm} \gammaJl \bigg( \int_\Omega \int_\Omega \tchins(\bx,\bRJ) \denM(\bx,\by) \tchins(\by,\bRJ) \dx \dy \bigg)  \nonumber \\
& & - \,4 \, \sum_J \sum_{lm} \gammaJl \, \sum_{J'} \int_\Omega \int_\Omega \pchis(\bx,\bRJp) {\big(\be - \bRJp \big)\mkern-4mu}_\beta \grada D(\bx,\by) \tchins(\by,\bRJ) \dx \dy \,, \label{Eq:sigmaIII} \\
\sigma \mkern-2mu ^{IV}_{\alpha \beta} & = & 
\frac{1}{4 \pi} \int_\Omega \grada \phi(\be,\bR) \gradb \phi(\be,\bR) \, \mathrm{d \be} + \sum_I \int_\Omega \grada b_I(\be,\bR_I) {\big(\be - \bR_I \big)\mkern-4mu}_\beta \Big(\phi(\be,\bR) - \frac{1}{2} \, V_I(\be,\bR_I) \Big) \, \mathrm{d \be}\nonumber \\
&-& \frac{1}{2} \sum_I \int_\Omega \grada V_I(\be,\bR_I) {\big(\be - \bR_I \big)\mkern-4mu}_\beta b_I(\be,\bR_I) \, \mathrm{d \be} + \frac{1}{2} \delta_{\alpha \beta} \int_\Omega \big(b(\be,\bR) - \rho_{\mathcal{D}}(\be) \big) \phi(\be,\bR) \, \mathrm{d \be} -\delta_{\alpha \beta} E_{\rm self}(\bR) \nonumber \\
&+ &  \sigma \mkern-2mu ^{E_c}_{\alpha\beta} \,.
\label{Eq:sigmaIV}
\end{eqnarray}
In the above equations, $\grada$ is the $\alpha^{th}$ component of the gradient vector $\gradx$; $\delta_{\alpha \beta}$ is the Kronecker delta function; $\sigma \mkern-2mu ^{E_c}_{\alpha\beta}$ is the stress tensor correction corresponding to overlapping pseudocharges \cite{sharma2018calculation}; the summation index $J$ runs over all atoms in $\Omega$; $J'$ runs over the $J^{th}$ atom and its periodic images; and  $I$ extends over all atoms in $\R^3$.  Note that, upon substitution of Eq.~\ref{Eqn:DensityMatrixDef} into the above expressions for the energy, atomic force, and stress tensor, the corresponding expressions in terms of Kohn-Sham orbitals \cite{ghosh2017sparc1, ghosh2017sparc2, sharma2018calculation} are readily obtained.

%%%%%%%%%%%%%%%%%%%%%%%%%%%%%%%%%%%%%%%%%%%%%%%%%%%%%%%%%%%%%%%%%%%%%%%%%%%%%%%%%%%%%%%%%%%%%%%%%%%%%%%%%%%%%%%%%%%%%%%%%%%%%%%%%%%%%%%%%%%%%%%%%%%%%%%%%%%%%%%%%%%%%%%%%%%%%%%%%
%%%%%%%%%%%%%%%%%%%%%%%%%%%%%%%%%%%%%%%%%%%%%%%%%%%%%%%%%%%%%%%%%%%%%%%%%%%%%%%%%%%%%%%%%%%%%%%%%%%%%%%%%%%%%%%%%%%%%%%%%%%%%%%%%%%%%%%%%%%%%%%%%%%%%%%%%%%%%%%%%%%%%%%%%%%%%%%%%
\section{Real-space DFT: density kernel formulation} \label{Sec:GS:Ds}
In this section, we formulate the electron density, electronic free energy, Hellmann-Feynman force, and Hellmann-Feynman stress tensor in terms of the density kernel \cite{ONETEP}, which in the current context 
%can be viewed as 
is the  matrix $\mathbf{D_s} = (D^s_{ij})$ corresponding to the single-particle density operator $\denM$ expressed in an orthonormal auxiliary orbitals basis $\{ \varphi_i(\bx) \}_{i=1}^{N_s}$, i.e., $D^s_{ij} = \bra{\varphi_i} \denM \ket{\varphi_j}$. In particular, 
\begin{equation}
\mathcal{D} = \sum_{i=1}^{N_s}\sum_{j=1}^{N_s}  \Ket{\varphi_i} D^s_{ij} \Bra{\varphi_j} \,, \quad \mathcal{D}(\bx,\by) = \sum_{i=1}^{N_s}\sum_{j=1}^{N_s} \varphi_i(\bx) D^s_{ij} \varphi_j(\by) \,, \label{Eqn:DensityOperatorDef:Aux}
\end{equation}
which corresponds to a unitary transformation of the Kohn-Sham orbitals
\begin{equation}
    \psi_i(\bx) = \sum_{j = 1}^{N_s} \varphi_j(\bx) Q_{ji} \,, \label{Eqn:SubspaceRotation}
\end{equation}
where $\mathbf{Q} = (Q_{ij})$ is the orthogonal matrix that diagonalizes $\mathbf{D_s}$:
\begin{equation}
     \mathbf{Q}^{\rm T} \mathbf{D_s} \mathbf{Q} = \mathrm{diag}\left(g_1,g_2,\ldots,g_{N_s}\right)\,.
\end{equation}
Hence, the auxiliary orbitals $\{ \varphi_i(\bx) \}_{i=1}^{N_s}$ span the same subspace as the Kohn-Sham orbitals $\{ \psi_i(\bx) \}_{i=1}^{N_s}$.

Let the subspace Hamiltonian $\mathbf{H_s} = (H^s_{ij})$  be the matrix representation of $\mathcal{H}$ in the orthonormal basis $\{ \varphi_i(\bx) \}_{i=1}^{N_s}$, i.e., $H^s_{ij} = \bra{\varphi_i} \mathcal{H} \ket{\varphi_j}$. It follows that the density kernel can be expressed in terms  of $\mathbf{H_s}$   as
\begin{eqnarray}
\mathbf{D_s} & = & g(\mathbf{H_s},\mu,\sigma) = \left( \mathbf{I} + \exp \left(\frac{\mathbf{H_s}-\mu \mathbf{I}}{\sigma} \right) \right)^{-1} \,, \label{Eqn:DsHsRelation} 
\end{eqnarray} 
where $\mathbf{I}$ is the identity matrix, and the Fermi level $\mu$ is determined from the constraint on the total number of electrons (Eq.~\ref{Eqn:TraceD}), which can be written in terms of the density kernel as:
\begin{equation}
    2\text{tr}(\mathbf{D_s}) \, = N_e\,, \label{Eqn:traceDs}
\end{equation}
where $\text{tr}(\cdot)$ denotes the trace. In arriving at this equation from Eq.~\ref{Eqn:TraceD}, we have used the relation
\begin{equation}\label{Eq:EquivalenceTrace}
\text{Tr}(\mathcal{D}) = \text{Tr} \left(\sum_{i=1}^{N_s}\sum_{j=1}^{N_s}  \Ket{\varphi_i} D^s_{ij} \Bra{\varphi_j} \right) = \sum_{i=1}^{N_s}\sum_{j=1}^{N_s}  \text{Tr} \left(\Ket{\varphi_i} D^s_{ij} \Bra{\varphi_j} \right) = \sum_{i=1}^{N_s}\sum_{j=1}^{N_s}  D^s_{ij} \Braket{ \varphi_j | \varphi_i }= \text{tr}(\mathbf{D_s}) \,.
\end{equation}

To make the expressions for electron density, atomic force, and stress analogous to those based on Kohn-Sham orbitals in the SPARC electronic structure code \cite{xu2021sparc, ghosh2017sparc2, ghosh2017sparc1}, into which we implement the proposed scheme, we introduce the density kernel transformed auxiliary orbitals $\{\widetilde{\varphi}_i(\bx)\}_{i=1}^{N_s}$:
\begin{equation}
    \widetilde{\varphi}_i(\bx) = \sum_{j = 1}^{N_s} \varphi_j(\bx) D^s_{ji} \,. \label{Eqn:phi_tilde}
\end{equation}
whereby, the density operator in Eq.~\ref{Eqn:DensityOperatorDef:Aux} takes the form
\begin{equation}
\mathcal{D} = \sum_{i = 1}^{N_s} \Ket{\widetilde{\varphi}_i} 
\Bra{\varphi_i}  \,, \quad \mathcal{D}(\bx,\by) = \sum_{i = 1}^{N_s} \widetilde{\varphi}_i(\bx) \varphi_i(\by) \,. \label{Eqn:DensityMatrix:NewForm}
\end{equation}
Thereafter, the electron density in Eq.~\ref{Eqn:rhoDensMat} can be written as
\begin{eqnarray}\label{Eqn:density:NewForm}
\rho_{\mathcal{D}}(\bx) = 2 \sum_{i=1}^{N_s} \widetilde{\varphi}_i(\bx) \varphi_i(\bx) \,.
\end{eqnarray}
Once the electronic ground state has been determined, the electronic free energy can be written as
\begin{eqnarray} \label{Eqn:FreeEnergy:NewForm}
\mathcal{F}(\bR) & = & 2\text{tr}(\mathbf{D_s H_s})   + E_{xc} \left[\rho_{\mathcal{D}}\right] - \int_{\Omega} V_{xc}\left[\rho_{\mathcal{D}}(\bx)\right]\rho_{\mathcal{D}}(\bx)\, \mathrm{d\bx} \nonumber + \frac{1}{2} \int_{\Omega}\big(b(\bx,\bR)-\rho_{\mathcal{D}}(\bx)\big) \phi(\bx,\bR) \, \mathrm{d\bx}  \nonumber  \\ 
& - & E_{\rm self}(\bR) + E_c(\bR) + 2 \sigma \text{tr}\left( \mathbf{D_s} \log \mathbf{D_s} + (\mathbf{I}-\mathbf{D_s}) \log (\mathbf{I}-\mathbf{D_s}) \right)  \,,
\end{eqnarray}
where the first and last terms, i.e., the band structure ($E_{band}$) and electronic entropy ($E_{ent}$) energies, have been obtained using orthogonality and trace relations as in Eq.~\ref{Eq:EquivalenceTrace}.

The nonlocal component of the atomic force in Eq.~\ref{Eqn:ON:forces} can be rewritten as
\begin{eqnarray}\label{Eqn:force_nl:NewForm}
\mathbf{f}^{nl}_I 
& = &  - 4 \text{Tr}\left(V_{nl,I} \nabla \mathcal{D} \right) \nonumber \\
& = &  -4 \text{Tr}\left(V_{nl,I} \nabla \sum_{i = 1}^{N_s} \Ket{\widetilde{\varphi}_i} \Bra{\varphi_i} \right) \nonumber \\
& = & -4  \sum_{i = 1}^{N_s} \Bra{\varphi_i} V_{nl,I} \Ket{\nabla \widetilde{\varphi}_i}   \nonumber \\
& = & -4  \sum_{i = 1}^{N_s} \Bra{\varphi_i} \sum_{lm} \gamma_{Il} \Ket{\tilde{\chi}_{Ilm}} \Braket{\tilde{\chi}_{Ilm} | \nabla \widetilde{\varphi}_i }  \nonumber \\
& = & -4 \sum_{i=1}^{N_s}  \sum_{lm} \gamma_{Il} 
 \bigg( \int_{\Omega} \varphi_{i}(\bx) \tilde{\chi}_{Ilm}(\bx,\bR_{I}) \, \mathrm{d\bx} \bigg) \, \bigg(\int_{\Omega} \nabla \widetilde{\varphi}_{i}(\bx) \tilde{\chi}_{Ilm}(\bx,\bR_{I}) \, \mathrm{d\bx} \bigg) \,     \,.
\end{eqnarray} 
Thereafter, the total Hellmann-Feynman atomic forces take the form
\begin{eqnarray}\label{Eqn:force:NewForm}
\mathbf{f}_I & = &  \sum_{I'} \int_{\Omega} \nabla b_{I'}(\bx,\bR_{I'}) \phi(\bx,\bR) \, \mathrm{d\bx} +  \mathbf{f}_{sc,I}  \nonumber \\
& - & 4 \sum_{i=1}^{N_s}  \sum_{lm} \gamma_{Il}  \bigg( \int_{\Omega} \varphi_{i}(\bx) \tilde{\chi}_{Ilm}(\bx,\bR_{I}) \, \mathrm{d\bx} \bigg) \, \bigg(\int_{\Omega} \nabla \widetilde{\varphi}_{i}(\bx) \tilde{\chi}_{Ilm}(\bx,\bR_{I}) \, \mathrm{d\bx} \bigg) \,.
\end{eqnarray} 

The stress tensor contributions arising from the electronic kinetic energy and nonlocal pseudopotential energy in Eqs.~\ref{Eq:sigmaI} and \ref{Eq:sigmaIII}, respectively, can be rewritten as
\begin{eqnarray}
\sigma \mkern-2mu ^{I}_{\alpha \beta} 
& = & 2 \int_\Omega  \bigg( \gradya \gradyb \mathcal{D}(\by,\be)\bigg)\bigg|_{\by = \be} \dx  \nonumber \\
& = & 2 \int_\Omega  \bigg( \gradya \gradyb \sum_{i = 1}^{N_s} \widetilde{\varphi}_i(\by) \varphi_i(\bx) \bigg)\bigg|_{\by = \be} \dx \nonumber \\
& = & 2 \sum_{i=1}^{N_s} \int_\Omega \varphi_i(\be) \grada \gradb \widetilde{\varphi}_i(\be) \dx \nonumber \\ 
& = &   -2 \sum_{i=1}^{N_s} \int_\Omega \grada \varphi_i(\be) \gradb \widetilde{\varphi}_i(\be) \dx  \,,
\label{Eq:stressIRe}
\end{eqnarray}
and
\begin{eqnarray}
\sigma \mkern-2mu ^{III}_{\alpha \beta} 
& = & -2 \delta_{\alpha \beta}  \sum_J \sum_{lm} \gammaJl \bigg( \int_\Omega \int_\Omega \tchins(\be,\bRJ) \mathcal{D}(\be, \by) \tchins(\by,\bRJ) \dx \dy \bigg) \nonumber \\
& &  -  4 \sum_J \sum_{lm} \gammaJl  \sum_{J'} \int_\Omega \int_\Omega \pchis(\be,\bRJp) {\big(\be - \bRJp \big)\mkern-4mu}_\beta \grada \mathcal{D}(\be, \by) \tchins(\by,\bRJ) \dx  \dy  \nonumber \\
& = & - 2 \delta_{\alpha \beta}  \sum_J \sum_{lm} \gammaJl \bigg( \int_\Omega \int_\Omega \tchins(\be,\bRJ) \sum_{i = 1}^{N_s} \widetilde{\varphi}_i(\bx) \varphi_i(\by) \tchins(\by,\bRJ) \dx \dy \bigg) \nonumber \\
& & - 4 \sum_J \sum_{lm} \gammaJl \sum_{J'} \int_\Omega \int_\Omega \pchis(\be,\bRJp) {\big(\be - \bRJp \big)\mkern-4mu}_\beta \grada \sum_{i = 1}^{N_s} \widetilde{\varphi}_i(\bx) \varphi_i(\by) \tchins(\by,\bRJ) \dx  \dy  \nonumber \\
& = & -2\delta_{\alpha \beta}  \sum_{i=1}^{N_s} \sum_J \sum_{lm} \gammaJl \bigg( \int_\Omega \tchins(\be,\bRJ) \widetilde{\varphi}_i(\be)  \dx \bigg) \bigg(\int_\Omega \tchins(\by,\bRJ) \varphi_i(\by) \dy \bigg) \nonumber \\
& & -  4 \sum_{i=1}^{N_s} \sum_J \sum_{lm} \gammaJl \Bigg(\sum_{J'} \int_\Omega \pchis(\be,\bRJp) {\big(\be - \bRJp \big)\mkern-4mu}_\beta \grada \widetilde{\varphi}_i(\be)  \dx \Bigg)\Bigg(\int_\Omega \tchins(\by,\bRJ) \varphi_i(\by) \dy\Bigg) \,.
\label{Eq:stressIIIRe}
\end{eqnarray}
Thereafter, the Hellmann-Feynman stress tensor takes the form 
\begin{eqnarray}
\sigma \mkern-2mu_{\alpha \beta} 
&=& \frac{1}{|\Omega |} \Bigg[  -2 \sum_{i=1}^{N_s} \int_\Omega \grada \varphi_i(\be) \gradb \widetilde{\varphi}_i(\be) \dx 
+ \delta_{\alpha \beta} \bigg( E_{xc}[\rho_{\mathcal{D}}] - \int_\Omega V_{xc}\big[\rho_{\mathcal{D}}(\bx)\big] \rho_{\mathcal{D}}(\bx) \dx \bigg) \nonumber \\
& - &  2 \delta_{\alpha \beta} \sum_{i=1}^{N_s} \sum_J \sum_{lm} \gammaJl \bigg( \int_\Omega \tchins(\be,\bRJ) \widetilde{\varphi}_i(\be)  \dx \bigg) \bigg(\int_\Omega \tchins(\by,\bRJ) \varphi_i(\by) \dy \bigg)  \nonumber \\
& - & 4 \sum_{i=1}^{N_s} \sum_J \sum_{lm} \gammaJl \Bigg(\sum_{J'} \int_\Omega \pchis(\be,\bRJp) {\big(\be - \bRJp \big)\mkern-4mu}_\beta \grada \widetilde{\varphi}_i(\be)  \dx \Bigg)\Bigg(\int_\Omega \tchins(\by,\bRJ) \varphi_i(\by) \dy\Bigg) \nonumber \\
&+& \frac{1}{4 \pi} \int_\Omega \grada \phi(\be,\bR) \gradb \phi(\be,\bR) \, \mathrm{d \be} + \sum_I \int_\Omega \grada b_I(\be,\bR_I) {\big(\be - \bR_I \big)\mkern-4mu}_\beta \Big(\phi(\be,\bR) - \frac{1}{2} \, V_I(\be,\bR_I) \Big) \, \mathrm{d \be}\nonumber \\
&-& \frac{1}{2} \sum_I \int_\Omega \grada V_I(\be,\bR_I) {\big(\be - \bR_I \big)\mkern-4mu}_\beta b_I(\be,\bR_I) \, \mathrm{d \be} + \frac{1}{2} \delta_{\alpha \beta} \int_\Omega \big(b(\be,\bR) - \rho_{\mathcal{D}}(\be) \big) \phi(\be,\bR) \, \mathrm{d \be} -\delta_{\alpha \beta} E_{\rm self}(\bR) \nonumber \\
&+ &  \sigma \mkern-2mu ^{E_c}_{\alpha\beta}  \Bigg] \,. \label{Eqn:Stress:NewForm}
\end{eqnarray}

%%%%%%%%%%%%%%%%%%%%%%%%%%%%%%%%%%%%%%%%%%%%%%%%%%%%%%%%%%%%%%%%%%%%%%%%%%%%%%%%%%%%%%%%%%%%%%%%%%%%%%%%%%%%%%%%%%%%%%%%%%%%%%%%%%%%%%%%%%%%%%%%%%%%%%%%%%%%%%%%%%%%%%%%%%%%%%%%%
%%%%%%%%%%%%%%%%%%%%%%%%%%%%%%%%%%%%%%%%%%%%%%%%%%%%%%%%%%%%%%%%%%%%%%%%%%%%%%%%%%%%%%%%%%%%%%%%%%%%%%%%%%%%%%%%%%%%%%%%%%%%%%%%%%%%%%%%%%%%%%%%%%%%%%%%%%%%%%%%%%%%%%%%%%%%%%%%%

\section{SQ3 method: Formulation and implementation} \label{Sec:SQ3}
In this section, we describe the calculation of  the auxiliary orbitals $\{\varphi_i(\bx)\}_{i=1}^{N_s}$ and density kernel transformed auxiliary orbitals $\{\tilde{\varphi}_i(\bx)\}_{i=1}^{N_s}$ in each iteration of the self-consistent field (SCF) method \cite{Martin2004}, 
%--- widely used solution strategy for determining the electronic ground state of the Kohn-Sham problem --- 
using which the electron density, electronic free energy, Hellmann-Feynman atomic forces, and Hellmann-Feynman stress tensor can be computed as discussed above. We refer to the resulting method as SQ3, given  its inherent similarity in philosophy and formulation to the linear scaling Spectral Quadrature (SQ) method \cite{suryanarayana2018sqdft, pratapa2016spectral, suryanarayana2013spectral}, with notable differences that the SQ3 method does not assume the subspace Hamiltonian to be sparse and does not employ truncation in the calculation of the density kernel,  making the computational cost scale cubically with $N_s$ rather than linearly.

In the SQ3 method, we perform Chebyshev filtering \cite{zhou2006self, zhou2006parallel} on the auxiliary orbitals from the previous SCF iteration followed by orthonormalization to generate the auxiliary orbitals $\{\varphi_i(\bx)\}_{i=1}^{N_s}$ for the current SCF iteration. To calculate $\{\tilde{\varphi}_i(\bx)\}_{i=1}^{N_s}$, the density kernel $\mathbf{D_s}$ needs to be determined (Eq.~\ref{Eqn:phi_tilde}), which we recall can be written in terms of the subspace Hamiltonian $\mathbf{H_s}$ --- projection of the Hamiltonian $\mathcal{H}$ on to the basis $\{\varphi_i(\bx)\}_{i=1}^{N_s}$ ---  using the Fermi-Dirac function $g$ (Eq.~\ref{Eqn:DsHsRelation}):
\begin{eqnarray}
\mathbf{D_s} & = & g(\mathbf{H_s},\mu,\sigma) = \left( \mathbf{I} + \exp \left(\frac{\mathbf{H_s}-\mu \mathbf{I}}{\sigma} \right) \right)^{-1} \,. \label{Eqn:DsHsRelation1} 
\end{eqnarray} 
One possible strategy to determine $\mathbf{D_s}$ is to perform an eigendecomposition of $\mathbf{H_s}$, an approach that would also make the Kohn-Sham orbitals available. However, this  method not only scales cubically with $N_s$, but also has limited scalability in parallel computations, which limits the minimum time to solution that can be reached in DFT simulations \cite{xu2021sparc}.  To delay this bottleneck for calculations at high temperature, we perform a  Fermi Operator Expansion (FOE) of the density kernel in terms of Chebyshev polynomials \cite{goedecker1994efficient, goedecker1995tight}:
\begin{eqnarray} \label{Eqn:ChebyshevExpansion}
  \mathbf{D_s}   \approx \sideset{}{'} \sum_{j=0}^{n_{pl}} c_j(\mu, \sigma) T_j\left( \frac{\mathbf{H_s}-\chi\mathbf{I}}{\xi} \right) 
    = \sideset{}{'} \sum_{j=0}^{n_{pl}} c_j(\mu, \sigma) T_j\left( \hatHs \right),  %\quad x \in [a,b] \,,
\end{eqnarray}
where the prime on the summation indicates that the first term is halved, $n_{pl}$ is the degree of the expansion, $T_j$ denotes the Chebyshev polynomial of degree $j$, $\chi = (\lambda_{N_s} + \lambda_1)/2, \xi = (\lambda_{N_s}-\lambda_1)/2$, and $\hatHs = \left(\mathbf{H_s}-\chi \mathbf{I} \right)/\xi$ is the scaled and shifted subspace Hamiltonian whose spectrum lies in the interval $[-1,1]$. The coefficients $c_j$ in the Chebyshev expansion can be determined using the relation
\begin{equation}
    c_j(\mu, \sigma) = \frac{2}{\pi} \int_{-1}^1 \frac{g(\xi \lambda+\chi, \mu, \sigma) T_j(\lambda)}{\sqrt{1-\lambda^2}} \mathrm{d} \lambda \,. \label{Eqn:ChebyshevCoeff}
\end{equation}

Analogous to the Clenshaw-Curtis SQ method \cite{suryanarayana2018sqdft, pratapa2016spectral}, the $n^{th}$ column of the density kernel in the SQ3 method is written as
\begin{equation} \label{Eq:DMCFinal}
\mathbf{D_s}  \mathbf{e}_n \approx  \sideset{}{'} \sum_{j=0}^{n_{pl}} c_j \mathbf{t}^j_n \,,
\end{equation}
where $\mathbf{e}_n$ is the standard basis vector, and $\mathbf{t}^j_n$ is the $n^{th}$ column of $T_j( \hatHs)$, determined using the three term recurrence relation for Chebyshev polynomials:
\begin{eqnarray}
    \mathbf{t}^0_n &=& \mathbf{e}_n, \nonumber\\
    \mathbf{t}^1_n &=& \hatHs \mathbf{e}_n, \nonumber\\
    \mathbf{t}^{j+1}_n &=& 2 \hatHs \mathbf{t}^j_n - \mathbf{t}^{j-1}_n, \quad j = 1,2,\ldots,n_{pl}-1\,. \label{Eqn:ThreeTermRecurrenceRelation}
\end{eqnarray}
The coefficients $c_j$ in Eq.~\ref{Eq:DMCFinal}  are determined using Eq.~\ref{Eqn:ChebyshevCoeff}, with the Fermi level $\mu$ chosen such that the constraint on the number of electrons is satisfied (Eq.~\ref{Eqn:traceDs}):
\begin{equation}\label{Eq:FermiLevelCalc}
 2 \sum_{n=1}^{N_s} \sideset{}{'} \sum_{j=0}^{n_{pl}} c_j(\mu, \sigma) \mathbf{e}_n^{\rm T}  \mathbf{t}^j_n = N_e\,.
\end{equation}

Once the density kernel has been so calculated, $\{\varphi_n(\bx)\}_{n=1}^{N_s}$ are transformed by the density kernel to obtain $\{\widetilde{\varphi}_n(\bx)\}_{n=1}^{N_s}$ (Eqn.~\ref{Eqn:phi_tilde}). The electron density is then calculated using Eqn.~\ref{Eqn:density:NewForm} and the electronic free energy is evaluated using  Eq.~\ref{Eqn:FreeEnergy:NewForm}, while employing the following expressions for the band structure and electronic entropy energies:
\begin{eqnarray}
E_{band} & \approx &   2 \sum_{n=1}^{N_s} \sideset{}{'} \sum_{j=0}^{n_{pl}} d_j  \mathbf{e}_n^{\rm T} \mathbf{t}^j_n \,, \\
E_{ent} & \approx & 2 \sum_{n=1}^{N_s} \sideset{}{'} \sum_{j=0}^{n_{pl}} e_j  \mathbf{e}_n^{\rm T} \mathbf{t}^j_n \,, 
\end{eqnarray} 
where the coefficients are calculated using the relations
\begin{eqnarray}
d_j & =&  \frac{2}{\pi} \int_{-1}^1 \frac{ (\xi \lambda+\chi) g(\xi \lambda+\chi, \mu, \sigma) T_j(\lambda)}{\sqrt{1-\lambda^2}} \mathrm{d} \lambda \,, \\
e_j & =&  \frac{2}{\pi} \int_{-1}^1 \frac{  (g(\xi \lambda+\chi, \mu, \sigma) \log g(\xi \lambda+\chi, \mu, \sigma) + (1-g(\xi \lambda+\chi, \mu, \sigma)) \log (1-g(\xi \lambda+\chi, \mu, \sigma))  ) T_j(\lambda)}{\sqrt{1-\lambda^2}} \mathrm{d} \lambda \,.
\end{eqnarray}
In arriving at the above expressions, we have performed Chebyshev polynomial expansions for $E_{band}$ and $E_{ent}$, similar to that done for the density kernel (Eq.~\ref{Eqn:ChebyshevExpansion}). At the electronic ground state, the Hellmann-Feynman atomic forces and stress tensor are computed using Eqns.~\ref{Eqn:force:NewForm} and \ref{Eqn:Stress:NewForm}, respectively.

We have implemented the SQ3 method in the SPARC electronic structure code \cite{xu2021sparc, ghosh2017sparc2, ghosh2017sparc1}. In particular, we build upon the implementation of the CheFSI method \cite{zhou2006self, zhou2006parallel} 
%--- default diagonalization based solution strategy for the Kohn-Sham problem 
in SPARC, which shares a number of computational kernels with the SQ3 method. Specifically, the Chebyshev filtering, orthogonalization, and projection kernels are used in the SQ3 implementation. We calculate the density kernel $\mathbf{D_s}$ using the iteration in Eq.~\ref{Eqn:ThreeTermRecurrenceRelation}, while parallelizing computations over the different columns of the density kernel. Such a scheme restricts the communication to only that required for the calculation of the Fermi level (Eq.~\ref{Eq:FermiLevelCalc}), while the computationally intensive step (Eq.~\ref{Eqn:ThreeTermRecurrenceRelation}) is  free from any data transfer between processors. To calculate $\{\widetilde{\varphi}_n(\bx)\}_{n=1}^{N_s}$ from $\{\varphi_n(\bx)\}_{n=1}^{N_s}$ (Eqn.~\ref{Eqn:phi_tilde}), we perform a dense matrix-matrix multiplication using the parallel \texttt{PDGEMM} routine in ScaLAPACK \cite{blackford1997scalapack}.  In doing so, we first use the \texttt{PDGEMR2D} routine to redistribute the density kernel from column-wise distribution to two-dimensional block-cyclic distribution, as required by \texttt{PDGEMM}. 

The degree of the the Chebyshev polynomial $n_{pl}$ required for a given accuracy is dependent on the the smearing $\sigma$, spectral width  of the subspace Hamiltonian $\mathbf{H_s}$ (i.e., $2 \xi$), and the relative location of the Fermi level $\mu$ within the spectrum of $\mathbf{H_s}$ \cite{suryanarayana2013spectral}.  In particular, the value of $n_{pl}$ required decreases with increasing temperature and decreasing spectral width, a consequence of  the increased smoothness of the Fermi-Dirac function and smaller interval over which it must be evaluated. Since the target application for the SQ3 method is systems at high temperature, and the spectral width of $\mathbf{H_s}$ is only slightly more than that of the occupied spectrum, low polynomial orders generally suffice so that the SQ3 method can be highly efficient, as we demonstrate below. And since computations at each real-space grid point are largely independent, the SQ3 method can attain excellent parallel scaling, as also demonstrated below.  It is worth noting that since $\mathbf{H_s}$ is dense, the computational cost of the SQ3 method as described above has $\mathcal{O}(N_s^3)$ scaling, similar to that for eigendecomposition, but with a smaller prefactor that decreases with increasing temperature. Indeed, it is possible to achieve $\mathcal{O}(N_s)$ scaling, as in the SQ method, if the auxiliary orbitals are such that $\mathbf{H_s}$ is sparse and truncation is adopted based on the decay of the density kernel. 

% %%%%%%%%%%%%%%%%%%%%%%%%%%%%%%%%%%%%%%%%%%%%%%%%%%%%%%%%%%%%%%%%%%%%%%%%%%%%%%%%%%%%%%%%%%%%%%%%%%%%%%%%%%%%%%%%%%%%%%%%%%%%%%%%%%%%%%%%%%%%%%%%%%%%%%%%%%%%%%%%%%%%%%%%%%%%%%%%%%
% %%%%%%%%%%%%%%%%%%%%%%%%%%%%%%%%%%%%%%%%%%%%%%%%%%%%%%%%%%%%%%%%%%%%%%%%%%%%%%%%%%%%%%%%%%%%%%%%%%%%%%%%%%%%%%%%%%%%%%%%%%%%%%%%%%%%%%%%%%%%%%%%%%%%%%%%%%%%%%%%%%%%%%%%%%%%%%%%%%
% %%%%%%%%%%%%%%%%%%%%%%%%%%%%%%%%%%%%%%%%%%%%%%%%%%%%%%%%%%%%%%%%%%%%%%%%%%%%%%%%%%%%%%%%%%%%%%%%%%%%%%%%%%%%%%%%%%%%%%%%%%%%%%%%%%%%%%%%%%%%%%%%%%%%%%%%%%%%%%%%%%%%%%%%%%%%%%%%%%

\section{Results and discussion} \label{Sec:Results}
In this section, we verify the accuracy and efficiency of the SQ3 method for the calculation of the electronic free energy, Hellmann-Feynman atomic forces, and Hellmann-Feynman stress tensor by comparison to conventional diagonalization based methods. For this purpose, we have implemented the SQ3 method in the SPARC electronic structure code \cite{xu2021sparc,  ghosh2017sparc2, ghosh2017sparc1}. We consider aluminum at density $2.7$ g/cc with temperatures ranging from $T=$ 10,000 K to 250,000 K.  We demonstrate the practical utility of the method by calculating the self-diffusion coefficient and viscosity of aluminum at temperature $T = 116,045$ K ($k_B T=10$ eV). In all simulations, we employ a twelfth-order accurate real-space finite-difference discretization, the local density approximation \cite{Kohn1965} for exchange-correlation interactions, a three-electron optimized norm conserving Vanderbilt (ONCV) pseudopotential \cite{hamann2013optimized} suitable for the range of temperatures considered, $\Gamma$-point Brillouin sampling, the restarted Periodic Pulay method \cite{banerjee2016periodic, pratapa2015restarted} for acceleration of the SCF iteration, and the alternating Anderson-Richardson (AAR) linear solver \cite{pratapa2016anderson,suryanarayana2019alternating} for  calculation of the electrostatic potential and application of the real-space Kerker preconditioner \cite{kumar2020preconditioning}.

%%%%%%%%%%%%%%%%%%%%%%%%%%%%%%%%%%%%%%%%%%%%%%%%%%%%%%%%%%%%%%%%%%%%%%%%%%%%%%%%%%%%%%%%%%%%%%%%%%%%%%%%%%%%%%%%%%%%%%%%%%%%%%%%%%%%%%%%%%%%%%%%%%%%%%%%%%%%%%%%%%%%%%%%%%%%%%%%%%%
\subsection{Accuracy and convergence} \label{Subsec:AccCon}
\begin{figure}[h!]
\centering
\subfloat[]{\includegraphics[keepaspectratio=true,width=0.4\textwidth]{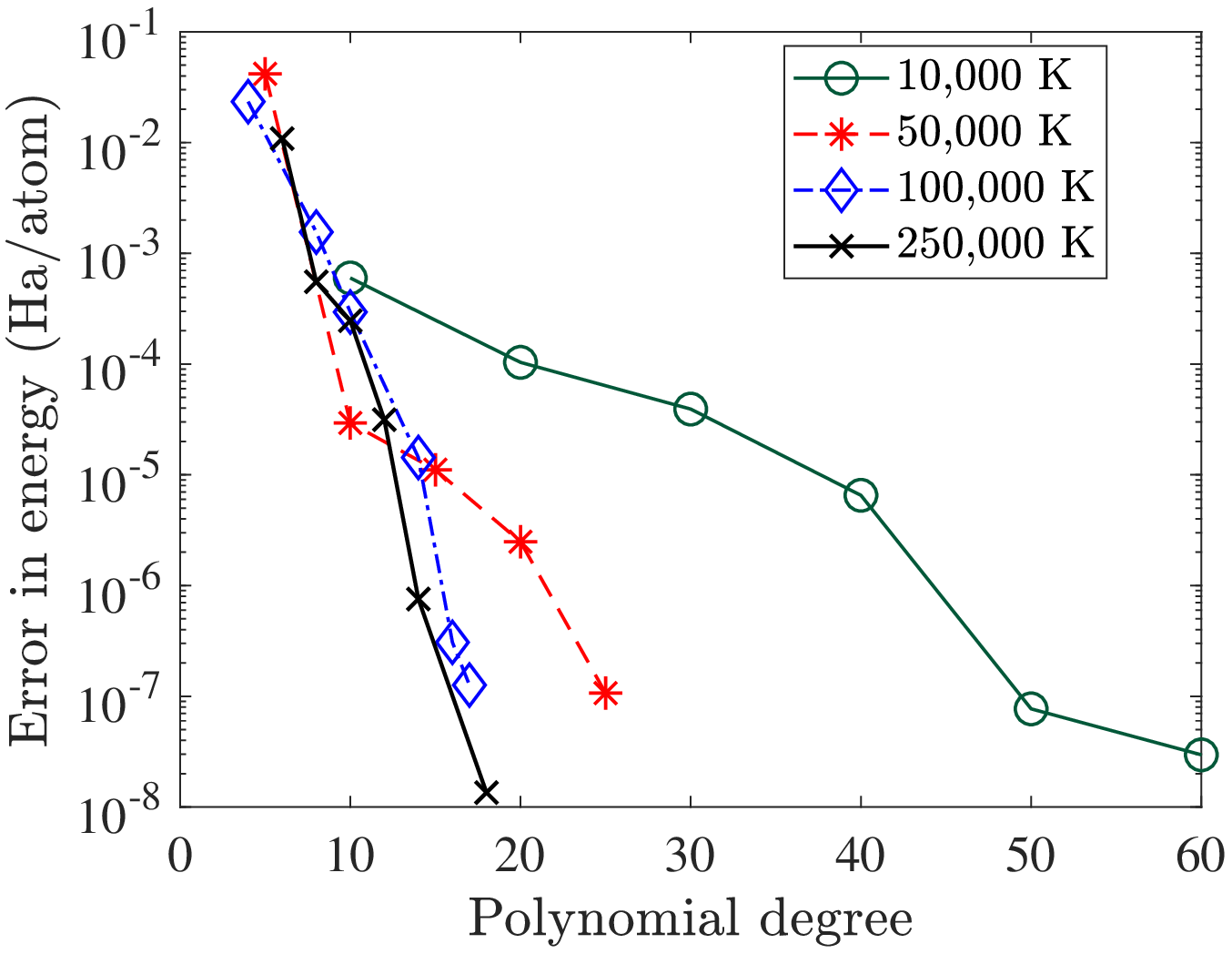} \label{Fig:EnergyErrorVsnpl} } 
\subfloat[]{\includegraphics[keepaspectratio=true,width=0.4\textwidth]{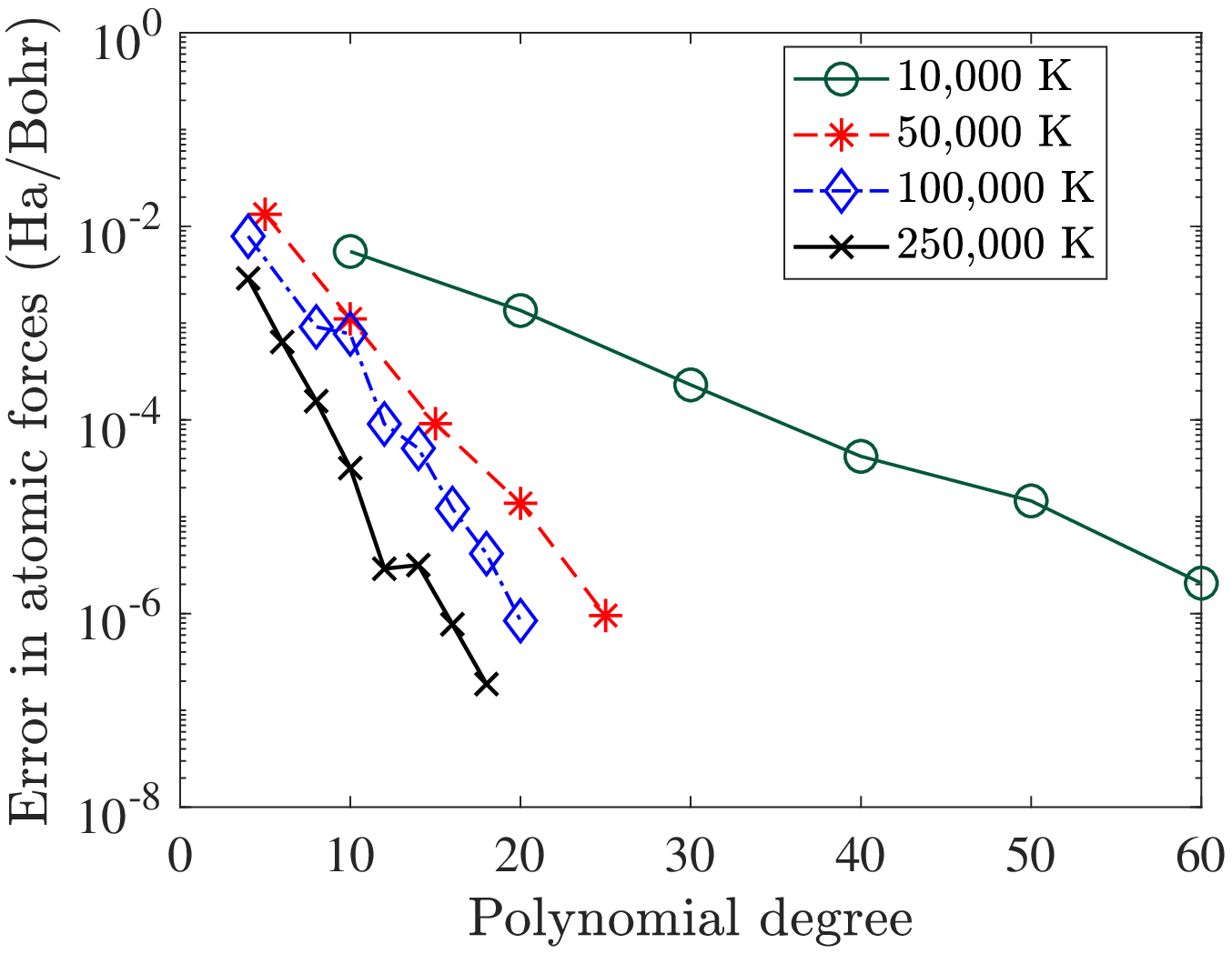} \label{Fig:ForceErrorVsnpl}} \\ 
\subfloat[]
{\includegraphics[keepaspectratio=true,width=0.4\textwidth]{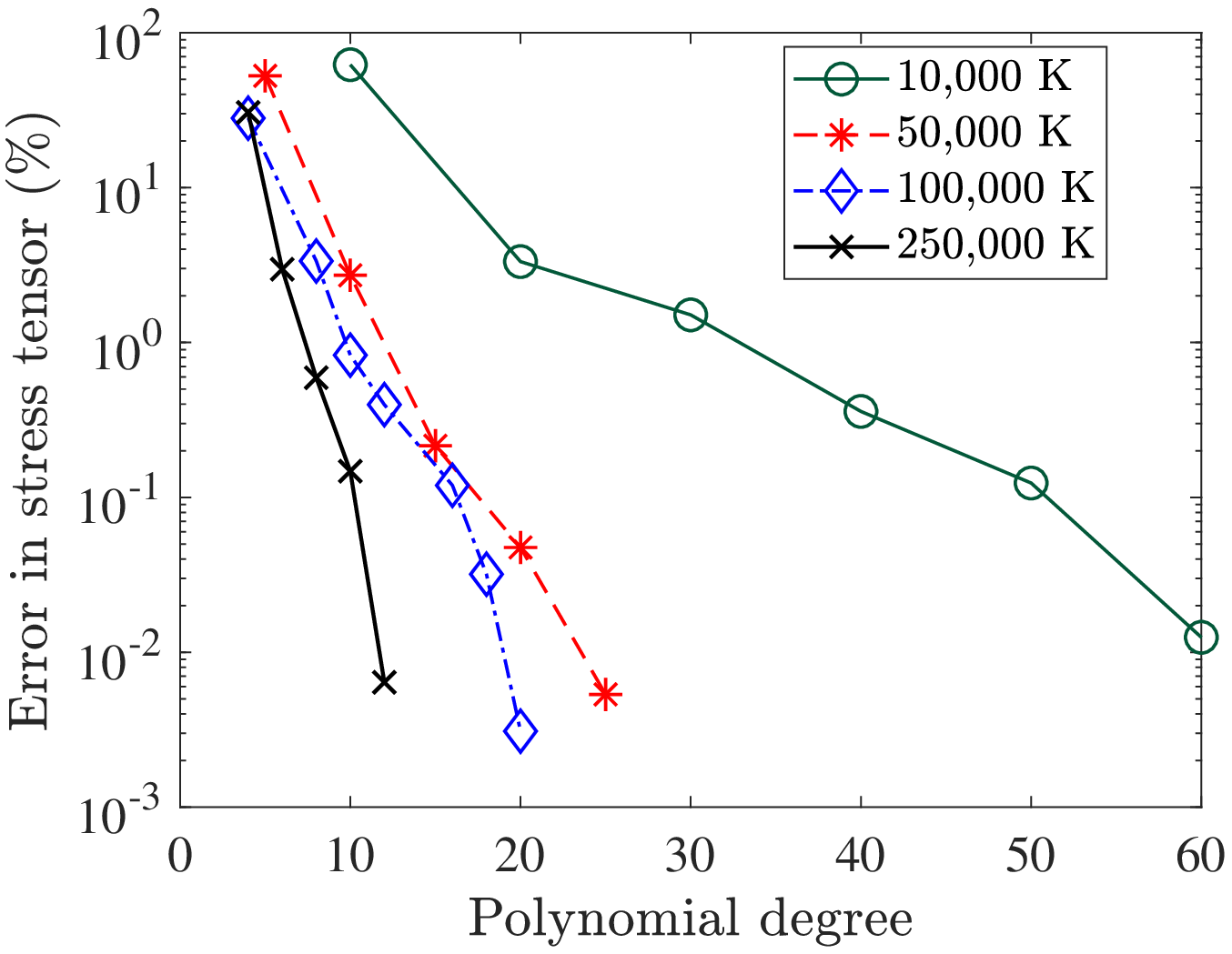} \label{Fig:StressErrorVsnpl}}
\caption{\label{Fig:AccConv:npl} Convergence of the electronic free energy, Hellmann-Feynman atomic forces, and Hellmann-Feynman stress tensor with respect to the polynomial degree $n_{pl}$ used in the Chebyshev polynomial expansion.  The error is defined to be the difference from the corresponding results obtained by diagonalization in SPARC.
}
\end{figure}

We first verify the accuracy and convergence of the SQ3 formulation and implementation by considering a $24$-atom aluminum cell at four different temperatures: $T=$ 10,000 K, 50,000 K, 100,000 K, and 250,000 K. In each system, all atoms are randomly perturbed by up to  $10\%$ of the nearest neighbor distance in an FCC configuration. We choose a mesh size of $0.5$ bohr to put associated discretization errors within chemical accuracy. In Fig.~\ref{Fig:AccConv:npl}, we plot the convergence of the electronic ground state free energy, Hellmann-Feynman atomic forces, and Hellmann-Feynman stress tensor with respect to $n_{pl}$, i.e., degree of polynomial used in the Chebyshev polynomial expansions for the density kernel, band structure energy, and electronic entropy energy.  The error is defined to be the difference from standard diagonalization results in the SPARC code. The results  show exponential convergence of the energy, atomic forces, and stress tensor with respect to  $n_{pl}$, with degree $\mathcal{O}(10-30)$ sufficient at the temperatures considered to attain accuracies typical in practice. The results also show that the polynomial degree required decreases with increasing temperature as expected.

%%%%%%%%%%%%%%%%%%%%%%%%%%%%%%%%%%%%%%%%%%%%%%%%%%%%%%%%%%%%%%%%%%%%%%%%%%%%%%%%%%%%%%%%%%%%%%%%%%%%%%%%%%%%%%%%%%%%%%%%%%%%%%%%%%%%%%%%%%%%%%%%%%%%%%%%%%%%%%%%%%%%%%%%%%%%%%%%%%%
\subsection{Performance and scaling}
We next investigate the efficiency of SQ3 relative to diagonalization for Kohn-Sham calculations at high temperature. For this purpose, we choose four FCC aluminum systems: (i) 1,568-atom cell at $T=$10,000 K, (ii) 500-atom cell at $T=$50,000 K, (iii) 256-atom cell at $T=$100,000 K, and (iv) 64-atom cell at $T=$250,000 K; all with $N_s=$ 10,000 and atoms randomly perturbed by up to $10\%$ of the nearest neighbor distance in the FCC configuration. The system sizes have been chosen to be comparable to those typical in QMD simulations at the associated temperatures.  In all cases, we select a mesh-size of $0.75$ bohr to put associated discretization errors within chemical accuracy, i.e., 0.001 Ha/atom, 0.001 Ha/bohr, and 1\% in the energy, force, and stress, respectively. In addition, we select $n_{pl} =$ 33, 12, 10, and 8 in SQ3 for the Al$_{1568}$, Al$_{500}$, Al$_{256}$, and Al$_{64}$ systems, respectively, to reduce $n_{pl}$ related errors to chemical accuracy also (Fig.~\ref{Fig:AccConv:npl}). All simulations have been performed on the \texttt{phoenix} cluster at Georgia Institute of Technology \cite{PhoenixClusterGT}.

In Table~\ref{tab:strong_scaling}, we compare the strong scaling performance of SQ3 and CheFSI  based diagonalization methods \cite{zhou2006self, zhou2006parallel}. In particular, we consider numbers of processors ranging from 24 to 1,000 and report the time taken per QMD step, along with the time taken for subspace diagonalization using ELPA \cite{marek2014elpa} and density kernel calculation in the diagonalization and SQ3 methods, respectively.  Indeed, these are the main computational kernels that are distinct between the two methods. We observe from the results that the SQ3 method scales up to $\mathcal{O}(1,000)$ processors, with further reduction in wall time possible for the density kernel calculation when more processors are utilized. The density kernel calculation does not scale ideally,  a consequence of the reduced effectiveness of the \texttt{BLAS3} operations as the number of columns of the density kernel associated with each processor becomes smaller. As expected, the speedup provided by the SQ3 method over diagonalization increases with increasing temperature, as the value of $n_{pl}$ required becomes smaller. In particular, while the minimum times to solution for diagonalization and SQ3 are similar for Al$_{1568}$ (10,000 K), SQ3 achieves an overall speedup of $\sim 1.3$, $\sim 1.5$, and $\sim 2.1$ for Al$_{500}$ Al$_{256}$, and Al$_{64}$ respectively, with corresponding speedups of the density kernel calculation over subspace diagonalization of  $\sim 2.5$, $\sim 2.7$, and $\sim 3.5$ respectively.

\begin{table}[h!]
\centering
%\resizebox{\textwidth}{!}{%
\begin{tabular}{|c||c|c||c|c||c|c||c|c||}
\hline
\multirow{2}{*}{\texttt{np}} & \multicolumn{2}{c||}{Al$_{1568}$, T = 10,000 K} & \multicolumn{2}{c||}{Al$_{500}$,   T = 50,000 K} & \multicolumn{2}{c||}{Al$_{256}$,   T = 100,000 K} & \multicolumn{2}{c||}{Al$_{64}$,   T = 250,000 K} \\ \cline{2-9} 
 & Diag & SQ3 & Diag & SQ3 & Diag & SQ3 & Diag & SQ3 \\ \hline
 
24 & \multicolumn{1}{c|}{1757.2 (84.8)} & 1970.1 (297.7) & \multicolumn{1}{c|}{580.5 (84.8)} & 598.7 (103.1) & \multicolumn{1}{c|}{383.5 (84.8)} & 382.1 (83.4) & \multicolumn{1}{c|}{173.6 (84.8)} & 151.4 (62.6) \\ \hline
48 & \multicolumn{1}{c|}{844.0 (49.8)} & 964.7 (170.5) & \multicolumn{1}{c|}{345.4 (49.8)} & 355.6 (60) & \multicolumn{1}{c|}{218.0 (49.8)} & 220.6 (52.4) & \multicolumn{1}{c|}{104.9 (49.8)} & 93.4 (38.3) \\ \hline
96 & \multicolumn{1}{c|}{518.6 (31.7)} & 568.6 (81.8) & \multicolumn{1}{c|}{192.7 (31.7)} & 188.0 (27.1) & \multicolumn{1}{c|}{124.7 (31.7)} & 116.5 (23.5) & \multicolumn{1}{c|}{64.1 (31.7)} & 51.1 (18.8) \\ \hline
192 & \multicolumn{1}{c|}{307.4 (27.0)} & 332.9 (52.6) & \multicolumn{1}{c|}{118.1 (27.0)} & 110.0 (18.9) & \multicolumn{1}{c|}{79.5 (27.0)} & 68.1 (15.6) & \multicolumn{1}{c|}{48.4 (27.0)} & 33.6 (12.3) \\ \hline
384 & \multicolumn{1}{c|}{224.2 (27.0)} & 236.0 (38.9) & \multicolumn{1}{c|}{81.6 (27.0)} & 68.0 (13.5) & \multicolumn{1}{c|}{59.7 (27.0)} & 44.8 (12.1) & \multicolumn{1}{c|}{40.2 (27.0)} & 22.5 (9.3) \\ \hline
500 & \multicolumn{1}{c|}{178.2 (27.0)} & 187.0 (35.8) & \multicolumn{1}{c|}{73.3 (27.0)} & 58.7 (12.4) & \multicolumn{1}{c|}{54.8 (27.0)} & 38.9 (11.1) & \multicolumn{1}{c|}{38.9 (27.0)} & 20.6 (8.8) \\ \hline
768 & \multicolumn{1}{c|}{170.7 (27.0)} & 176.2 (32.6) & \multicolumn{1}{c|}{67.1 (27.0)} & 51.5 (11.4) & \multicolumn{1}{c|}{52.5 (27.0)} & 35.8 (10.4) & \multicolumn{1}{c|}{38.8 (27.0)} & 20.3 (8.6) \\ \hline
1000 & \multicolumn{1}{c|}{173.4 (27.0)} & 177.2 (30.8) & \multicolumn{1}{c|}{61.8 (27.0)} & 45.5 (10.8) & \multicolumn{1}{c|}{49.7 (27.0)} & 32.6 (9.9) & \multicolumn{1}{c|}{37.5 (27.0)} & 18.2 (7.8) \\ \hline
\end{tabular}%
%}
\caption{Strong scaling comparison of CheFSI diagonalization and SQ3 methods. The timings are in seconds and correspond to the wall time for a single QMD step, with 6 SCF iterations required for convergence in all instances. The numbers in parentheses for diagonalization represent the wall times for diagonalization of the subspace Hamiltonian, performed using ELPA \cite{marek2014elpa}. The numbers in parentheses for SQ3 represent the wall times for the density kernel calculation. }
\label{tab:strong_scaling}
\end{table}

Overall, we find that the SQ3 method is able to delay the cubic scaling bottleneck for Kohn-Sham calculations at high temperature, with increasing advantages as the temperature and/or number of processors is increased. We note that we have verified that alternate forms of parallelization such as block cyclic decomposition of the $\hatHs$ and $T_j\left( \hatHs \right)$ matrices --- matrix-matrix multiplication routine replacing the iteration in Eq.~\ref{Eqn:ThreeTermRecurrenceRelation}, as in conventional FOE methods --- are not as efficient as the approach adopted here ($\sim 1.3$x slower). It is worth noting also that since the SQ3 method does not employ truncation and has cubic scaling with $N_s$, the linear scaling SQ method \cite{suryanarayana2018sqdft, pratapa2016spectral, suryanarayana2013spectral} would become the method of choice at temperatures higher than those considered here.

%%%%%%%%%%%%%%%%%%%%%%%%%%%%%%%%%%%%%%%%%%%%%%%%%%%%%%%%%%%%%%%%%%%%%%%%%%%%%%%%%%%%%%%%%%%%%%%%%%%%%%%%%%%%%%%%%%%%%%%%%%%%%%%%%%%%%%%%%%%%%%%%%%%%%%%%%%%%%%%%%%%%%%%%%%%%%%%%%%%
\subsection{High temperature QMD: self-diffusion coefficient and viscosity of aluminum at 116,045 K}

We now calculate the self-diffusion coefficient and viscosity of aluminum at density 2.7 g/cc and temperature 116,045 K ($k_B T = 10$ eV). Specifically, we consider a 108-atom unit cell in the isokinetic ensemble using a Gaussian thermostat \cite{NVK}. In order to efficiently obtain sufficient statistics, we average over QMD simulations corresponding to 20 different initial conditions for the atomic positions and velocities --- obtained by performing 20 orbital-free DFT  \cite{ghosh2016higher, suryanarayana2014augmented} simulations, each run for 40,000 steps with a time step of 0.15 fs --- where each simulation has been run for more than 24,000 steps with a time step of 0.15 fs. Since macroscopic dynamical properties can be written as time integrals of associated microscopic time correlation functions using Green–Kubo (GK) relations \cite{hansen2013theory}, we calculate the self-diffusion coefficient $D$ and viscosity $\eta$ using the expressions
\begin{eqnarray}
D(t) & = & \frac{1}{3N}\sum_{i=1}^N\int_0^t\langle \bm{v}_i(\tau)\cdot\bm{v}_i(0) \rangle \text{d}\tau \,, \\
\eta(t) & =& \frac{|\Omega|}{k_BT}\int_0^t\left(\frac{1}{5}\sum_{i=1}^5\langle s_i(\tau)s_i(0)\rangle\right)\text{d}\tau \,,
\end{eqnarray}
where $\langle . \rangle$ denotes the ensemble average, $\bm{v}_i$ is the velocity, and $s_i$ are the independent components of the deviatoric (i.e., traceless) stress tensor: $\sigma_{12}$, $\sigma_{23}$, $\sigma_{31}$,  $(\sigma_{11}-\sigma_{22})/2$, and $(\sigma_{22}-\sigma_{33})/2$ \cite{Alfe1998}. 

\begin{figure}[h!]
\centering
\subfloat[]{\includegraphics[keepaspectratio=true,,width=0.48\textwidth]{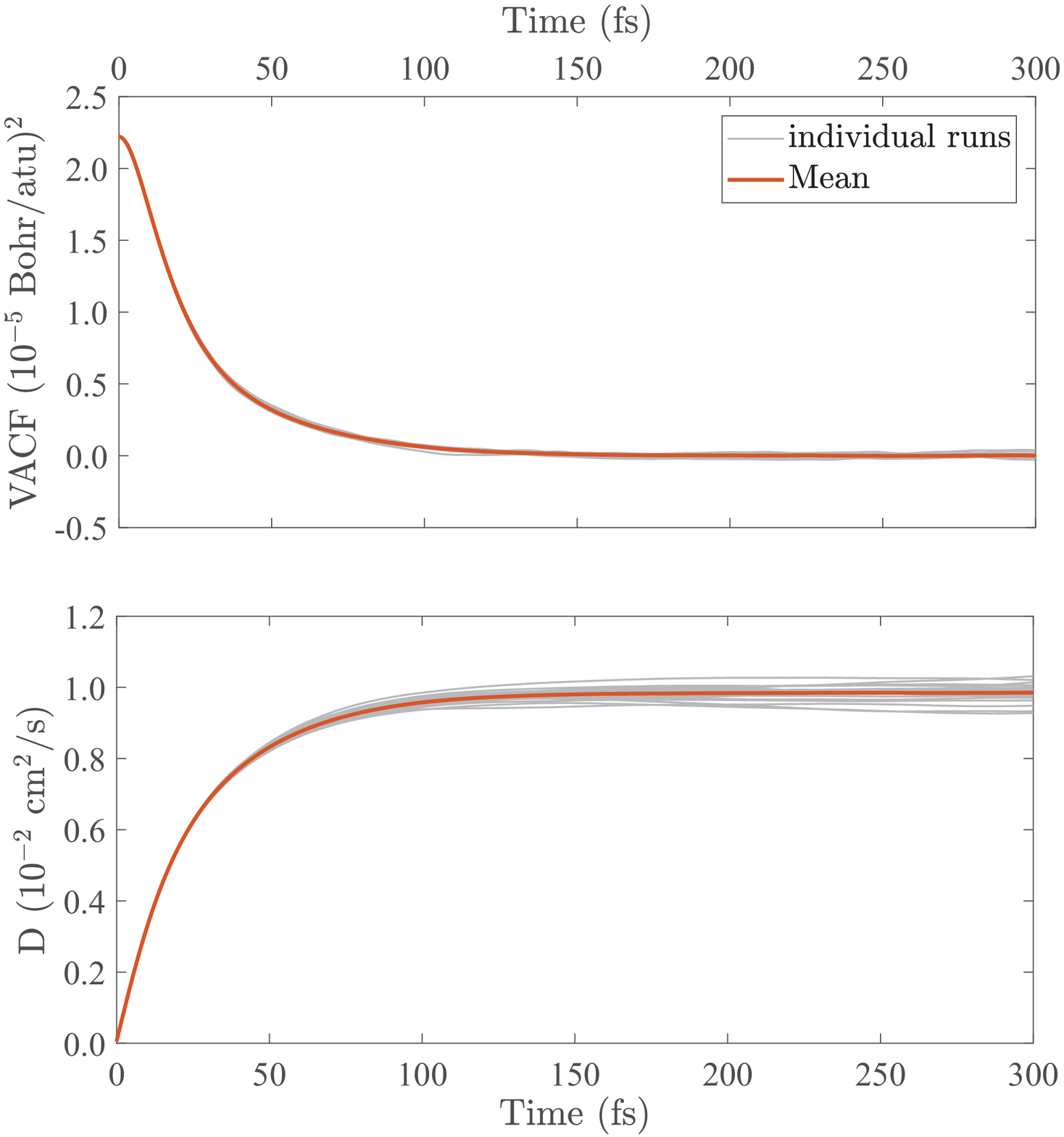} \label{Fig:DiffusionCoeff} } 
\subfloat[]{\includegraphics[keepaspectratio=true,,width=0.48\textwidth]{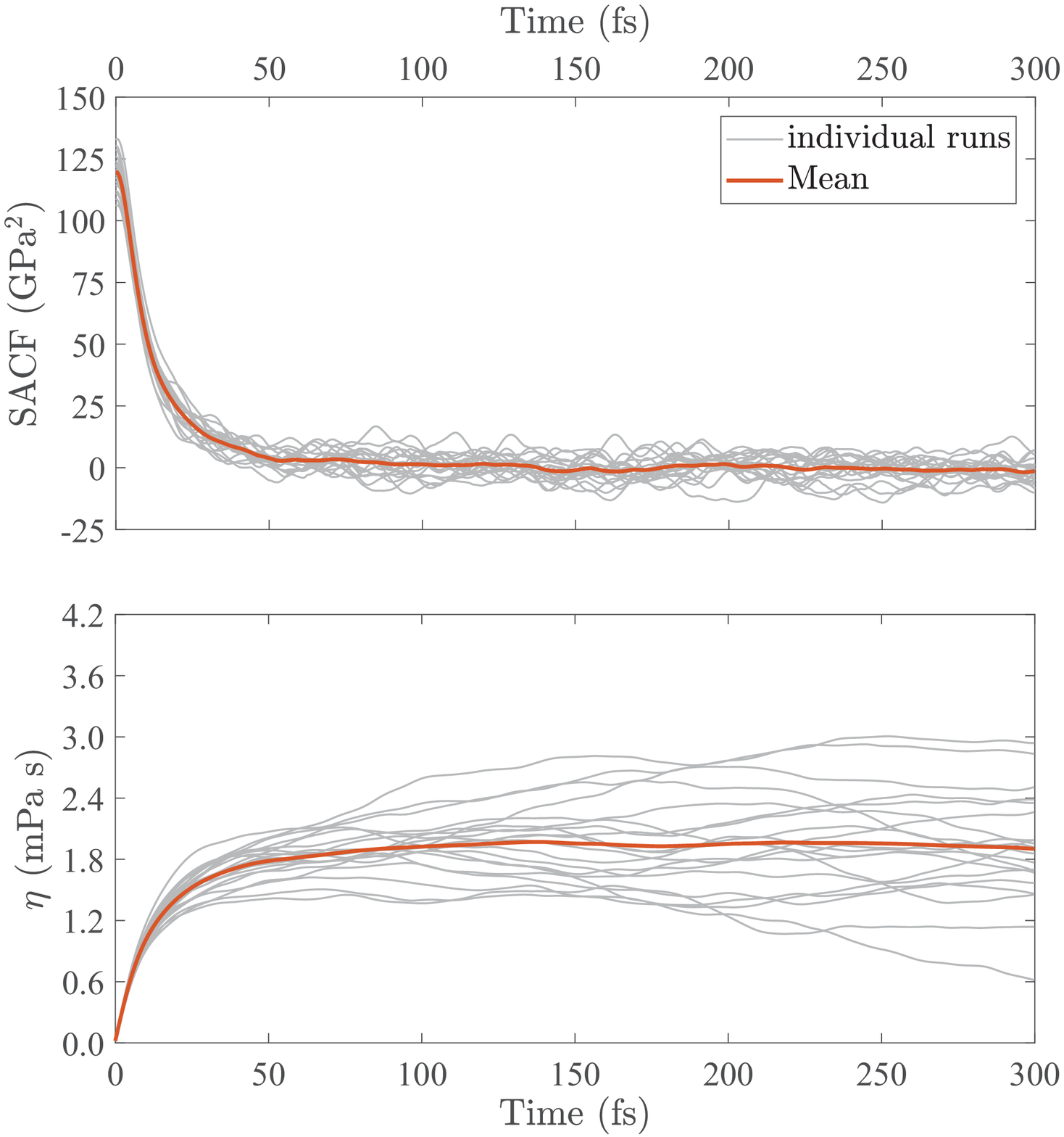} \label{Fig:Viscosity}} 
\caption{\label{Fig:QMD} (a) Velocity autocorrelation function (VACF) and self-diffusion coefficient, and (b) stress autocorrelation function (SACF) and viscosity for aluminum at  density 2.7 g/cc and temperature 116,045 K.}
\end{figure}

We present the results so obtained in Fig.~\ref{Fig:QMD}. It is clear that the both the velocity autocorrelation function (VACF) (i.e., $\frac{1}{N}\sum_{i=1}^N \langle \bm{v}_i(\tau)\cdot\bm{v}_i(0) \rangle$) and stress autocorrelation function (SACF) (i.e., $\frac{1}{5}\sum_{i=1}^5\langle s_i(\tau)s_i(0)\rangle$) decay in $\sim 100$ fs, yielding a self-diffusion coefficient and viscosity of (0.984 $\pm$ 0.001)$\times10^{-2}$ cm$^2$/s and 1.943 $\pm$ 0.015 mPa$\cdot$s, respectively. The present full Kohn–Sham DFT result for the self-diffusion coefficient is consistent with recent orbital-free DFT calculations \cite{sjostrom2015ionic}, where a value of 0.93 $\times 10^{-2}$ cm$^2$/s was reported, while being free of the kinetic and entropic energy approximations inherent in orbital-free DFT.

%%%%%%%%%%%%%%%%%%%%%%%%%%%%%%%%%%%%%%%%%%%%%%%%%%%%%%%%%%%%%%%%%%%%%%%%%%%%%%%%%%%%%%%%%%%%%%%%%%%%%%%%%%%%%%%%%%%%%%%%%%%%%%%%%%%%%%%%%%%%%%%%%%%%%%%%%%%%%%%%%%%%%%%%%%%%%%%%%%
%%%%%%%%%%%%%%%%%%%%%%%%%%%%%%%%%%%%%%%%%%%%%%%%%%%%%%%%%%%%%%%%%%%%%%%%%%%%%%%%%%%%%%%%%%%%%%%%%%%%%%%%%%%%%%%%%%%%%%%%%%%%%%%%%%%%%%%%%%%%%%%%%%%%%%%%%%%%%%%%%%%%%%%%%%%%%%%%%%
%%%%%%%%%%%%%%%%%%%%%%%%%%%%%%%%%%%%%%%%%%%%%%%%%%%%%%%%%%%%%%%%%%%%%%%%%%%%%%%%%%%%%%%%%%%%%%%%%%%%%%%%%%%%%%%%%%%%%%%%%%%%%%%%%%%%%%%%%%%%%%%%%%%%%%%%%%%%%%%%%%%%%%%%%%%%%%%%%%

\section{Concluding remarks} \label{Sec:Conclusions}
We presented the SQ3 method, a density matrix based method for Kohn-Sham calculations at high temperature that eliminates the need for diagonalization, thus reducing the cost of such calculations significantly relative to conventional diagonalization based approaches. We developed real-space expressions for the electron density, electronic free energy, Hellmann-Feynman forces, and Hellmann-Feynman stress tensor in terms of an orthonormal auxiliary orbital basis and its density kernel transform. Using Chebyshev filtering to generate the auxiliary basis, we then developed an approach akin to Clenshaw-Curtis spectral quadrature to compute the individual columns of the density kernel based on the Fermi operator expansion in Chebyshev polynomials; and employed a similar approach to evaluate band structure and entropic energy components. Upon implementation of the method in the SPARC electronic structure code  \cite{xu2021sparc, ghosh2017sparc2, ghosh2017sparc1}, we found systematic convergence to exact diagonalization results and significant speedups relative to conventional diagonalization based methods of up to $\sim 2$x, with increasing advantages as the temperature and/or number of processors is increased. Finally, we employed the new method to compute the self-diffusion coefficient and viscosity of aluminum at 116,045 K from Kohn-Sham quantum molecular dynamics, where we found agreement with previous more approximate orbital-free density functional methods.

The combination of conventional diagonalization based methods at low temperatures, SQ3 at moderately high temperatures (e.g., $\mathcal{O}$(10,000)--$\mathcal{O}$(250,000) kelvin), and SQ at higher temperatures enables accurate and efficient Kohn-Sham calculations over the full range of temperatures from ambient to millions of kelvin. The use of a localized orthonormal basis as in the discrete discontinuous basis projection (DDBP) method \cite{xu2018discrete} is likely to further increase the efficiency of such calculations, making it a worthy subject for future research.

%%%%%%%%%%%%%%%%%%%%%%%%%%%%%%%%%%%%%%%%%%%%%%%%%%%%%%%%%%%%%%%%%%%%%%%%%%%%%%%%%%%%%%%%%%%%%%%%%%%%%%%%%%%%%%%%%%%%%%%%%%%%%%%%%%%%%%%%%%%%%%%%%%%%%%%%%%%%%%%%%%%%%%%%%%%%%%%%%%
%%%%%%%%%%%%%%%%%%%%%%%%%%%%%%%%%%%%%%%%%%%%%%%%%%%%%%%%%%%%%%%%%%%%%%%%%%%%%%%%%%%%%%%%%%%%%%%%%%%%%%%%%%%%%%%%%%%%%%%%%%%%%%%%%%%%%%%%%%%%%%%%%%%%%%%%%%%%%%%%%%%%%%%%%%%%%%%%%%
%%%%%%%%%%%%%%%%%%%%%%%%%%%%%%%%%%%%%%%%%%%%%%%%%%%%%%%%%%%%%%%%%%%%%%%%%%%%%%%%%%%%%%%%%%%%%%%%%%%%%%%%%%%%%%%%%%%%%%%%%%%%%%%%%%%%%%%%%%%%%%%%%%%%%%%%%%%%%%%%%%%%%%%%%%%%%%%%%%

\section*{Acknowledgements}
This work was performed in part under the auspices of the U.S. Department of Energy by Lawrence Livermore National Laboratory under Contract DE-AC52-07NA27344. Q.X., X.J., B.Z., and P.S. gratefully acknowledge support from the U.S. Department of Energy, Office of Science under grant DE-SC0019410. J.E.P. gratefully acknowledges support from the ASC/PEM program at LLNL. The views and conclusions contained in this document are those of the authors and should not be interpreted as representing the official policies, either expressed or implied, of the Department of Energy, or the U.S. Government.
 
\section*{Data Availability}
The data that support the findings of this study are available from the corresponding author upon reasonable request.

\section*{Author Declarations} 
The authors have no conflicts to disclose.

%%%%%%%%%%%%%%%%%%%%%%%%%%%%%%%%%%%%%%%%%%%%%%%%%%%%%%%%%%%%%%%%%%%%%%%%%%%%%%%%%%%%%%%%%%%%%%%%%%%%%%%%%%%%%%%%%%%%%%%%%%%%%%%%%%%%%%%%%%%%%%%%%%%%%%%%%%%%%%%%%%%%%%%%%%%%%%%%%%
%%%%%%%%%%%%%%%%%%%%%%%%%%%%%%%%%%%%%%%%%%%%%%%%%%%%%%%%%%%%%%%%%%%%%%%%%%%%%%%%%%%%%%%%%%%%%%%%%%%%%%%%%%%%%%%%%%%%%%%%%%%%%%%%%%%%%%%%%%%%%%%%%%%%%%%%%%%%%%%%%%%%%%%%%%%%%%%%%%
%%%%%%%%%%%%%%%%%%%%%%%%%%%%%%%%%%%%%%%%%%%%%%%%%%%%%%%%%%%%%%%%%%%%%%%%%%%%%%%%%%%%%%%%%%%%%%%%%%%%%%%%%%%%%%%%%%%%%%%%%%%%%%%%%%%%%%%%%%%%%%%%%%%%%%%%%%%%%%%%%%%%%%%%%%%%%%%%%%

%\bibliography{bibliography}

\end{document}